\documentclass[11pt,a4paper]{article}
\usepackage[T1]{fontenc}
\usepackage{lmodern,amsmath,booktabs,tabularx,graphicx}
\usepackage[margin=2.65cm]{geometry}
\usepackage{xurl}
\usepackage[hidelinks]{hyperref}
\DeclareRobustCommand{\appname}{\url{https://lhc-bsm-excesses-explorer.app.cern.ch/}}
\title{Mapping LHC BSM Excesses --- From an Extended Higgs-Sector Survey to an Interactive Catalogue}
\author{Livia Soffi\thanks{Co-convener of the LHC Beyond the Standard Model Working Group.} \\ \small INFN, Sezione di Roma, Rome, Italy}
\date{26 November 2026}
\begin{document}
\maketitle
\begin{abstract}
This work began with a survey of searches for an extended Higgs sector at the Large Hadron Collider (LHC), presented by the author in an \href{https://indico.ijclab.in2p3.fr/event/12648/contributions/42756/}{invited plenary talk} at Higgs Hunting 2026, a Higgs physics workshop held at IJCLab in Orsay and APC in Paris. The survey mapped the hypotheses and mass ranges covered by public ATLAS and CMS searches and recorded the deviations reported in those searches. The author subsequently extended the approach across the LHC search programme for physics beyond the Standard Model (BSM), building a LHC BSM excess inventory and its public web application at \appname{}. It provides experimentalists, phenomenologists and other readers of LHC results with an interactive catalogue of reported excesses. The application gives an overview of the curated BSM entries, allows entries to be filtered by signature and other search parameters, and links each entry to the original public result and, where identified, to related searches. Its statistical quantities are taken from the public results. This paper documents the process that led from the extended Higgs-sector survey to the application, including the definition of the source catalogues, the verification and selection of reported excesses, the identification of counterpart searches, and the construction of the combined BSM workbook that drives the application.

\end{abstract}

\section{Motivation and sources}
The LHC Beyond the Standard Model Working Group (LHC BSM WG) brings together experimentalists and phenomenologists to develop common benchmarks and discuss the interpretation of LHC searches~\cite{wgmandate}. The Working Group maintains a public list of reported excesses for cross-experiment follow-up~\cite{wg}. The author contributed to the development of this list as a member of the Working Group's Steering Committee. As the range of signatures and datasets studied at the LHC grows, connecting results through a common signal hypothesis helps identify independent tests of a reported excess and complementary channels that probe the same physics. The benchmark and the parameter range covered by each search provide the basis for that comparison.

The extended Higgs-sector survey presented by the author at Higgs Hunting 2026 applied this approach to public ATLAS and CMS searches at the LHC~\cite{hh}. It mapped the tested hypotheses and mass ranges, recorded the reported deviations and identified searches relevant to the same benchmarks. The method was then recently extended to the wider LHC BSM search programme, covering a broader range of signatures and observables.

The author developed the LHC BSM excess inventory and the public application at \appname{}. The inventory is stored in the combined BSM workbook described in Sec.~\ref{sec:bsm-inventory}. The application makes those entries searchable and links them to their experimental references. The input catalogues are CMS Results Search~\cite{cmscatalog}, the ATLAS public-result catalogues for searches~\cite{atlassearch} and for Higgs and di-Higgs results~\cite{atlashiggs}, and the LHC BSM WG excess inventory~\cite{wg}. The collaboration papers and official public notes supply the reported masses and statistical quantities. The derived work consists of selecting, transcribing, standardising, linking and displaying the published information. Fields for unreported quantities are left empty.

\section{Method developed for the extended Higgs sector}
The author compiled the extended Higgs-sector workbook presented at Higgs Hunting 2026~\cite{hh} to document search coverage independently of whether a search reported a deviation. A search entry associates a public analysis reference with a process, reconstructed final state, dataset and scanned parameter range. A publication can contribute multiple entries when it tests distinct hypotheses or channels. The local significance \(Z_{\mathrm{local}}\) quantifies an upward fluctuation at a specified signal hypothesis in Gaussian standard deviations (\(\sigma\)). The review highlighted deviations with \(Z_{\mathrm{local}}>2.4\), and the extended Higgs-sector workbook retained all selected searches. The resulting survey comprised 179 search entries from 174 public references~\cite{hh}.

The construction separated record proposal from source validation. Codex assisted with locating public documents, drafting structured records and identifying candidate counterparts. Review against the original results resolved the identity of each mass coordinate, the statistical meaning of the quoted values and the scope of candidate matches. Accepted records supplied the inputs to the figure scripts.

\subsection{Projection and coverage}
The two figures discussed here illustrate selected parts of the extended Higgs-sector survey. The accompanying studies map mass coverage by search category and reconstructed final state, catalogue exotic decay signatures of the 125~GeV Higgs boson, and examine cross-experiment follow-up near 95~GeV and in the 650--700~GeV region. The study of the latter region also considers the production and spin hypotheses relevant to cross-experiment comparisons~\cite{hh}.

\begin{figure}[!htbp]
  \centering
  \includegraphics[width=.94\linewidth]{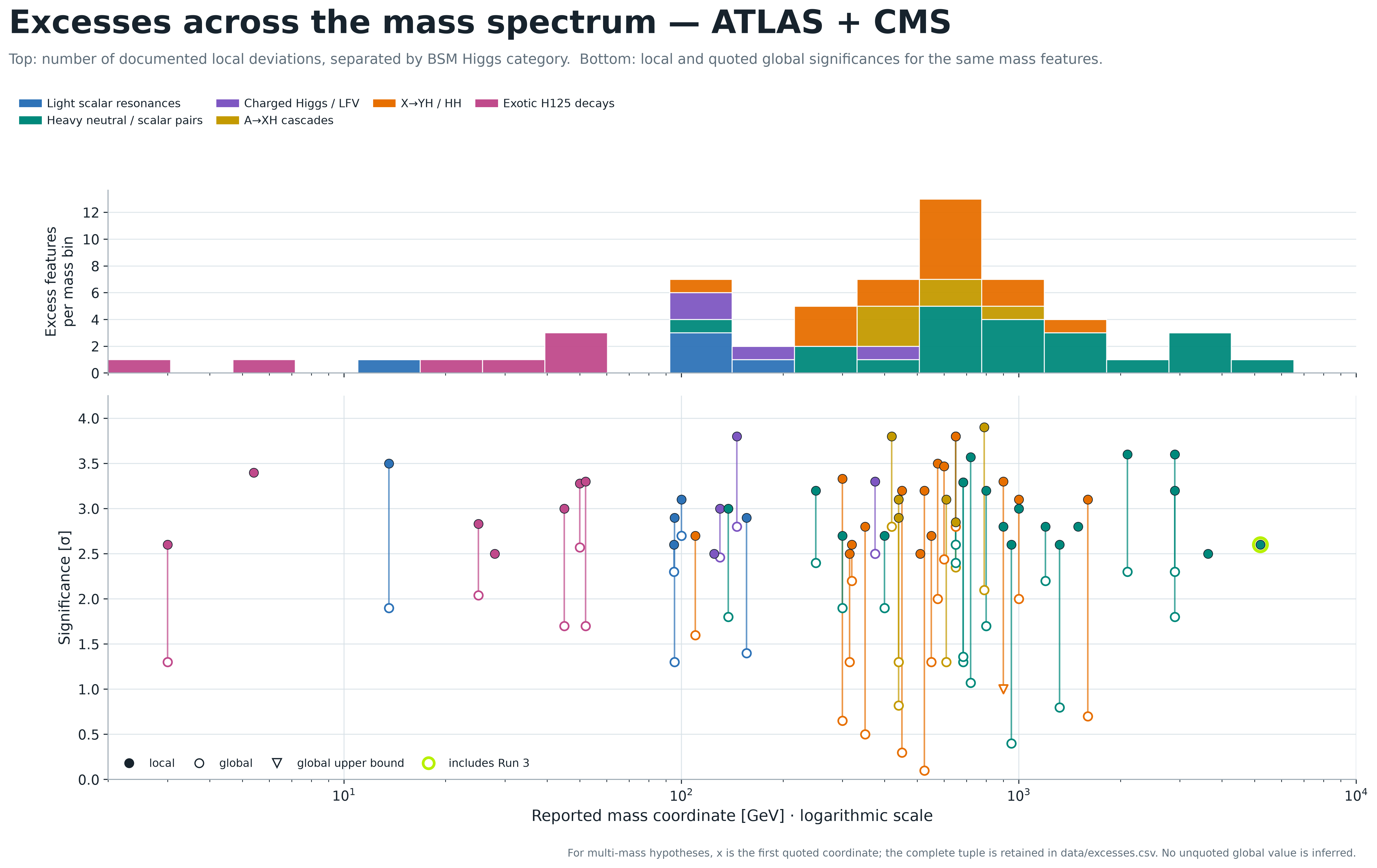}
  \caption{Extended Higgs-sector deviations shown on slide~7 of Ref.~\cite{hh}. The upper panel counts reported features by category. The lower panel pairs local and quoted global significances for each feature, using downward triangles for global upper bounds. The first quoted mass is used for multi-mass hypotheses.}
  \label{fig:hh26-spectrum}
\end{figure}

The first figure discussed below asks where the reported deviations occur in mass and what significances are quoted. The second asks which searches cover a particular cascade benchmark. The mass spectrum in Fig.~\ref{fig:hh26-spectrum} addresses the first question by placing the selected deviations on a common mass axis while preserving the distinction between local significances, quoted global significances and upper bounds.

\begin{figure}[!htbp]
  \centering
  \includegraphics[width=.94\linewidth]{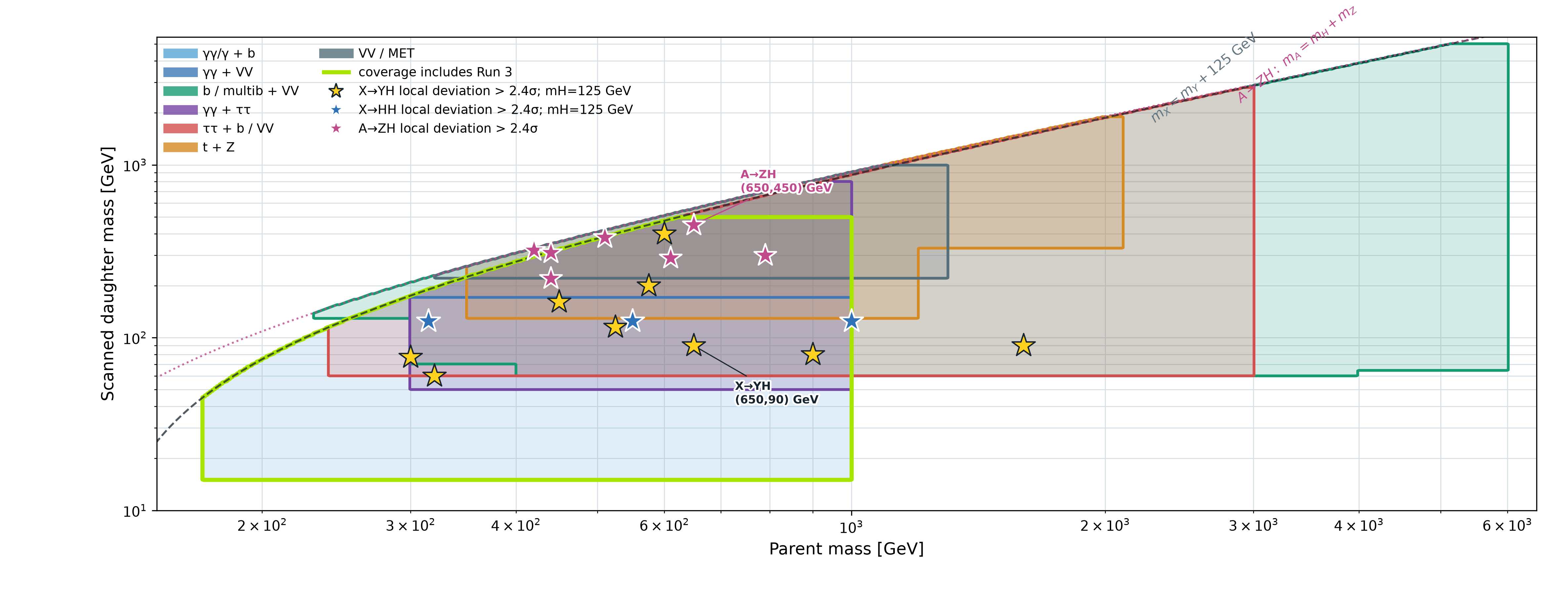}
  \caption{Published ATLAS and CMS cascade search coverage in the parent--daughter mass plane, from slide~18 of Ref.~\cite{hh}. Shading groups the tested final states and stars locate selected deviations.}
  \label{fig:hh26-cascades}
\end{figure}

Figure~\ref{fig:hh26-cascades} addresses the second question by retaining both mass coordinates. In \(X\to YH\), \(X\) denotes the parent and \(Y\) the scanned daughter produced with the 125~GeV Higgs boson \(H\). In \(A\to ZH\), \(A\) denotes the parent and \(H\) the scanned Higgs state produced with the \(Z\) boson. The benchmarks \((650,90)\) and \((650,450)\)~GeV in these respective searches share a parent mass and probe distinct hypotheses. The coverage map locates candidate searches. Benchmark compatibility and sensitivity are determined from the original analysis. This distinction motivates retaining the full mass tuple and the tested process in each workbook record.

\section{Construction of the LHC BSM excess inventory}
\label{sec:bsm-inventory}
The author constructed the LHC BSM excess inventory to make follow-up of reported deviations systematic across the full scope of the study. Starting from a reported excess, a reader needs to locate the original result, identify the tested signal hypothesis and the location of the fluctuation, and find independent searches that can test the same benchmark. The extended Higgs-sector survey illustrated this approach for mass-based benchmarks. Applying it to the wider programme calls for a common representation that also accommodates deviations reported in signal regions or classifier bins.

A publication can report several distinct deviations, each with its own benchmark and significance. The combined BSM workbook therefore uses the individual excess as its unit and records links to relevant searches.

The construction proceeds in five stages. First, the public catalogues define the source populations and each entry is resolved to a verified primary document. Second, full-text scans of the CMS and ATLAS documents identify candidate passages, and the linked LHCb papers are reviewed for the entries selected by the Working Group. Third, the reported significance and location of each candidate are validated against the source. Fourth, the validated records are represented as rows in experiment-specific workbooks and counterpart searches are reviewed. Finally, the workbooks and the approved counterpart links are merged into the versioned file served by the application.

\subsection{Experiment sources and document verification}
Catalogue construction precedes significance screening so that the source population is defined independently of its outcome. The CMS import used the BSM theme in the EXO, NPS, SUS, B2G and HIG groups~\cite{cmscatalog}. The ATLAS import merged 1,342 records from the searches and Higgs/di-Higgs catalogues into 1,038 unique records using document type and official reference~\cite{atlassearch,atlashiggs}. The LHCb component starts from the eight LHCb entries in the LHC BSM WG list and verifies them against the three linked papers~\cite{wg}. Table~\ref{tab:input-audit} distinguishes documents screened, distinct results in the application and excess rows for the 19 November 2026 source snapshot.

\begin{table}[!htbp]
\centering\small
\begin{tabularx}{\linewidth}{@{}lXrrr@{}}
\toprule
Experiment & Input documents by type & PDFs verified & Distinct results & Excess rows \\
\midrule
ATLAS & 814 papers, 78 conference notes, 146 public notes & \(1038/1038\) & 98 & 126 \\
CMS   & 643 papers, 303 physics analysis summaries & \(946/946\) & 124 & 145 \\
LHCb  & 3 papers linked from the WG list & \(3/3\) & 3 & 8 \\
\midrule
Total & & \(1987/1987\) & 225 & 279 \\
\bottomrule
\end{tabularx}
\caption{Source coverage and counts for the deployed BSM workbook. The input document types follow the catalogue metadata. The distinct-result and excess-row columns count public analysis references and individual workbook entries, respectively.}
\label{tab:input-audit}
\end{table}

The document-identity audit checked 1,987 PDFs. Document identity was checked against the official identifier or the opening-page title, with manual verification where needed. The source URL and audit outcome were retained for each file. A paper takes precedence over a preliminary note for the same result and dataset. Results with distinct data exposures can remain separate despite a common signal hypothesis.

\subsection{Selection and representation}
The CMS and ATLAS scans use the complete PDF text because a local excess may be reported in the body, a table or a figure caption. Text normalisation handles PDF line breaks and typographic variants. The candidate searches recognise \(\sigma\), standard-deviation wording, \emph{s.d.} and explicit local/global pairs. The ATLAS retrieval takes the union of passages extracted independently with Poppler and Ghostscript and searches for fluctuation language near significance or \(p\)-value expressions. The CMS candidate-validation run reviewed 299 records, while the broader ATLAS retrieval flagged 807 of its 1,038 input documents. The two counts describe the workloads of experiment-specific retrieval rules.

Validation checks the source passage and, where relevant, the corresponding figure or table. The common criterion for the CMS and ATLAS scans is a current observed upward fluctuation under a BSM signal hypothesis, with an explicitly quoted local significance \(Z_{\mathrm{local}}\geq2.0\). Curated exceptions are recorded separately. Expected sensitivity, deficits, quoted results from other analyses and deviations in ordinary Standard Model Higgs measurements are excluded. Searches for exotic Higgs decays can qualify under the same BSM requirement.

Review decisions and source passages are recorded for ambiguous cases. Distinct fluctuations receive separate rows. Alternative interpretations of the same fluctuation share a row, and neighbouring fits in a correlated mass scan require review before they can be counted separately. The workbook retains the public reference, final state, dataset, publication year, significance values and the status of any quoted global value. Typed locations distinguish mass points, tuples and intervals from thresholds, signal regions and classifier bins. Global upper bounds retain their status, and \(p\)-values are transcribed as significances only where the source supplies the conversion.

A coverage check traced the 54 CMS and 27 ATLAS references in the existing LHC BSM WG control list~\cite{wg} through the source catalogues and the combined BSM workbook. The primary documents, including their tables and figures, were used to verify the correspondence between those references and the workbook entries. The LHCb entries preserve the distinction between a pointwise deviation of the observed limit and a quoted local significance.

\subsection{Counterpart matching}
The CMS--ATLAS counterpart review for results published from 2024 onwards searches the complete opposite-experiment catalogue, including results outside the excess selection. Titles, abstracts and keywords retrieve candidate analyses. Acceptance requires comparison of the primary documents for signal topology, final state and parameter coverage. Dataset and luminosity provide context. The most recent analysis is preferred when several results provide equivalent tests.

A \emph{direct} match tests the same signal in an equivalent channel and compatible parameter range. \emph{Partial equivalence} specifies the overlapping part of the tested scope. A \emph{complementary channel} addresses the same broader hypothesis through a different signature. For each approved relation, the combined BSM workbook records the counterpart reference and a separate \emph{Counterpart rationale} field. The rationale identifies the shared production or decay hypothesis, the relevant final state and the overlap in parameter coverage. It also records differences that determine the extent of the match. This is particularly useful when a single analysis covers several signal models: readers can see which component supports the relation and check the judgement against the linked public documents. When both analyses enter the inventory, the same reviewed relation can be followed in both directions. Each reported significance remains attached to its original benchmark and analysis.

\subsection{Versioned implementation}
The private CERN GitLab repository stores the import and validation scripts, catalogue snapshots, document manifests, review decisions, experiment workbooks and application code. The primary PDFs are kept in the local audit corpus. The repository records their public source links and the evidence used for validation. Each experiment-specific builder combines the validated records with catalogue metadata and writes a comma-separated values (CSV) file with a shared schema. The merge script checks that schema, retains the scientific fields, assigns a sequential identifier to each row and attaches approved counterpart links by analysis reference. Tests check catalogue coverage, workbook structure and application rendering. Codex assisted in drafting and testing extraction scripts, locating passages missed by initial text searches, and proposing structured records and counterpart candidates. Candidate passages were checked for document identity, observed versus expected context, and the association between a significance and its signal benchmark. The author established the physics criteria and adjudicated contested values and matches against the source documents. Reviewed decisions are stored separately from generated outputs, so the workbook can be rebuilt from the versioned records and deterministic scripts.

\section{Application and worked example}
The public application at \appname{}~\cite{explorer} provides two views of the combined BSM workbook. The plots provide an overview of the catalogue and of a selected subset. The table supports targeted reading by analysis reference, experiment, dataset, final state, publication year, mass where applicable, and reported significance. The analysis links take the reader to the original public results. The counterpart columns identify analyses to consult when following up a signal hypothesis. The \emph{Counterpart rationale} column, available through \emph{Choose columns}, shows the scientific basis for each recorded match.

\begin{figure}[!b]
  \centering
  \includegraphics[width=\linewidth]{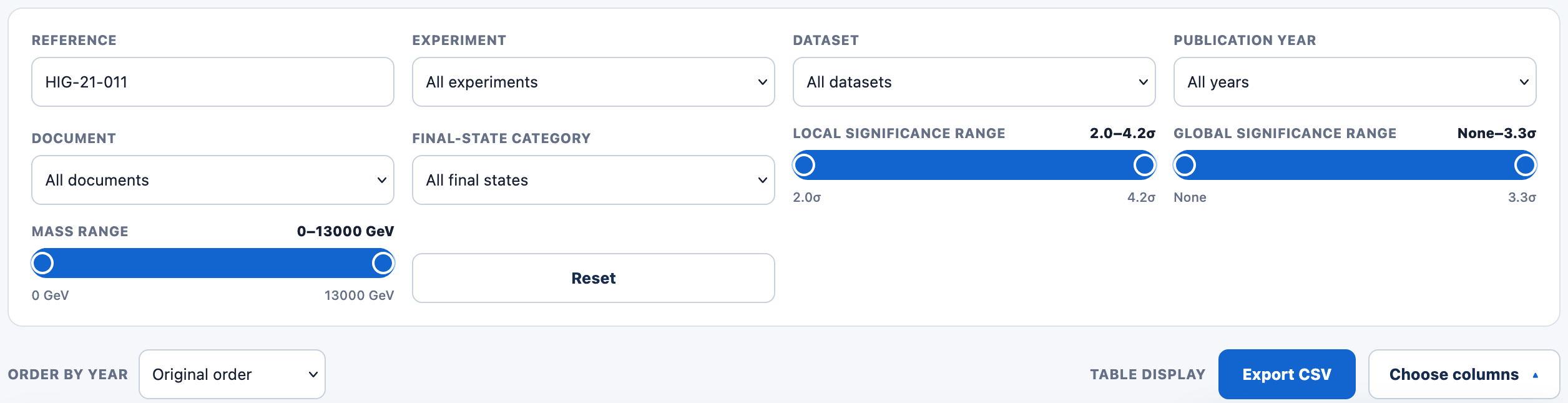}
  \caption{Table filters and display controls in the application, with the reference field set to \texttt{HIG-21-011}.}
  \label{fig:explorer-filters}
\end{figure}

A read-only Flask service on the CERN Platform as a Service (PaaS)~\cite{cernpaas} loads the versioned workbook and joins public titles and document types from catalogue metadata. The table and plots share the filters shown in Fig.~\ref{fig:explorer-filters}. Selected rows can be exported as CSV, and plots as PNG or PDF with the applied selection. The application presents and links the curated records. The experimental meaning of each quantity is established in the corresponding primary document. The counts and figures refer to the inventory based on the source snapshot of 19 November 2026.

\begin{figure}[p]
  \centering
  \includegraphics[width=.98\linewidth,trim=0 0 0 665pt,clip]{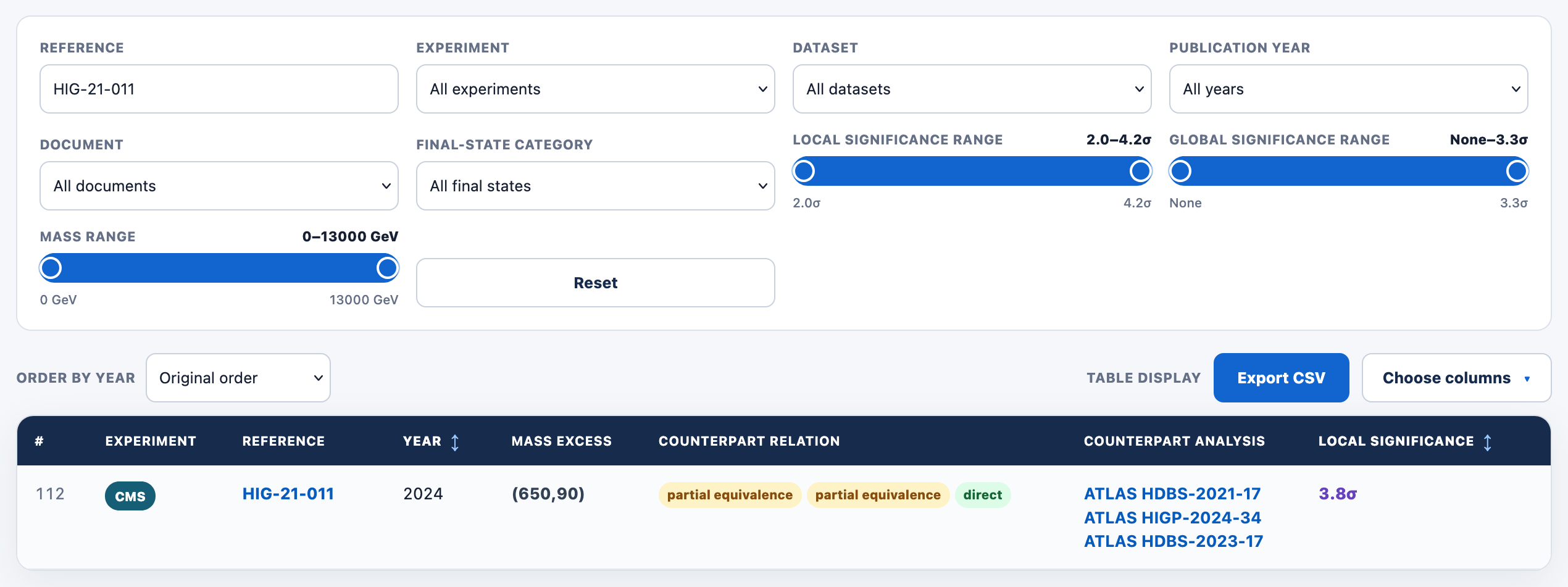}
  \caption{Reference search for \texttt{HIG-21-011} in the application at \appname{}. The record preserves both masses and links to the source and the reviewed ATLAS counterparts. CMS quotes \(3.8\sigma\) local and a global significance below \(2.8\sigma\) at this point~\cite{cmscascade}.}
  \label{fig:explorer-table}

  \smallskip
  \includegraphics[width=.81\linewidth]{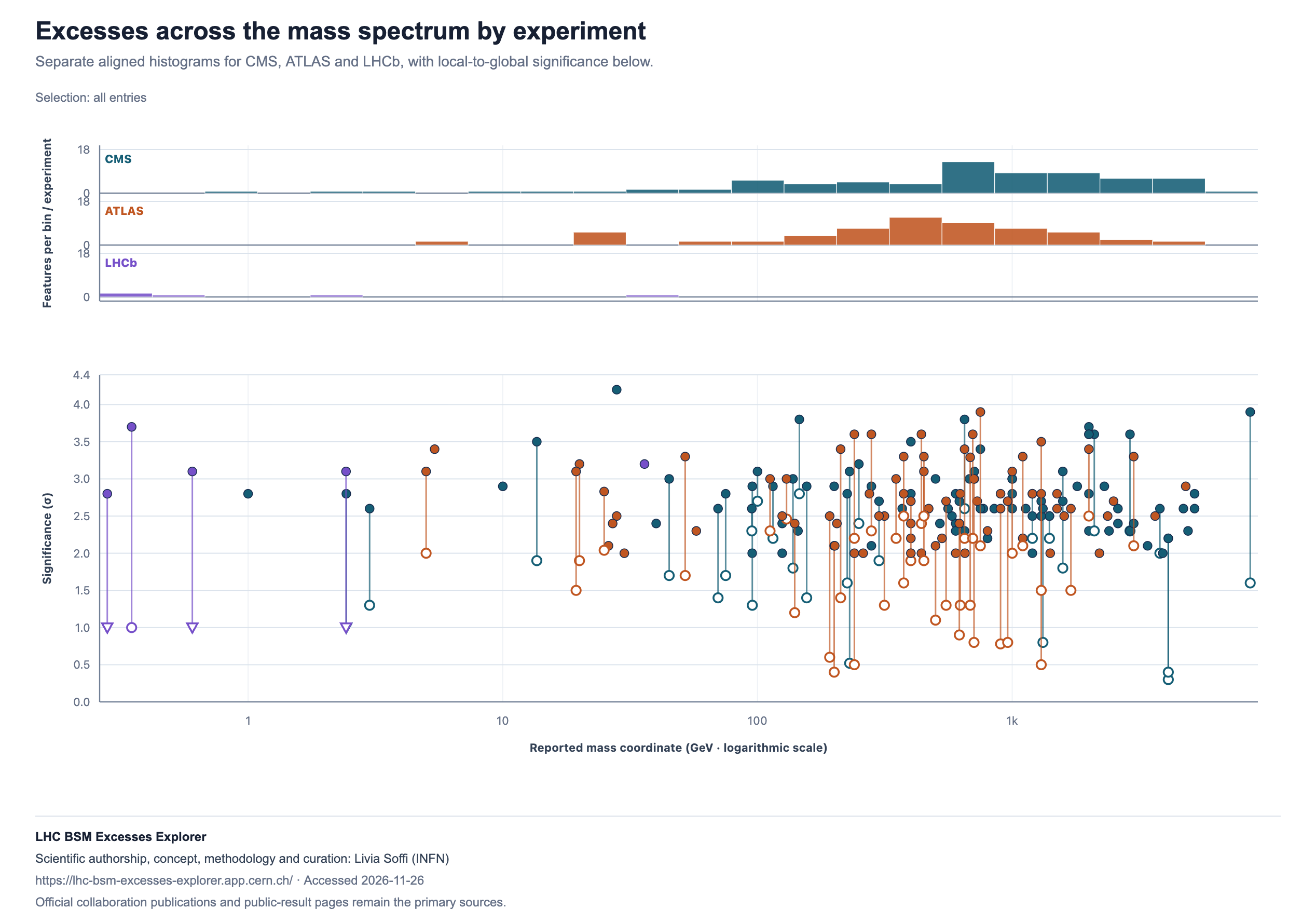}\par\smallskip
  \includegraphics[width=.81\linewidth]{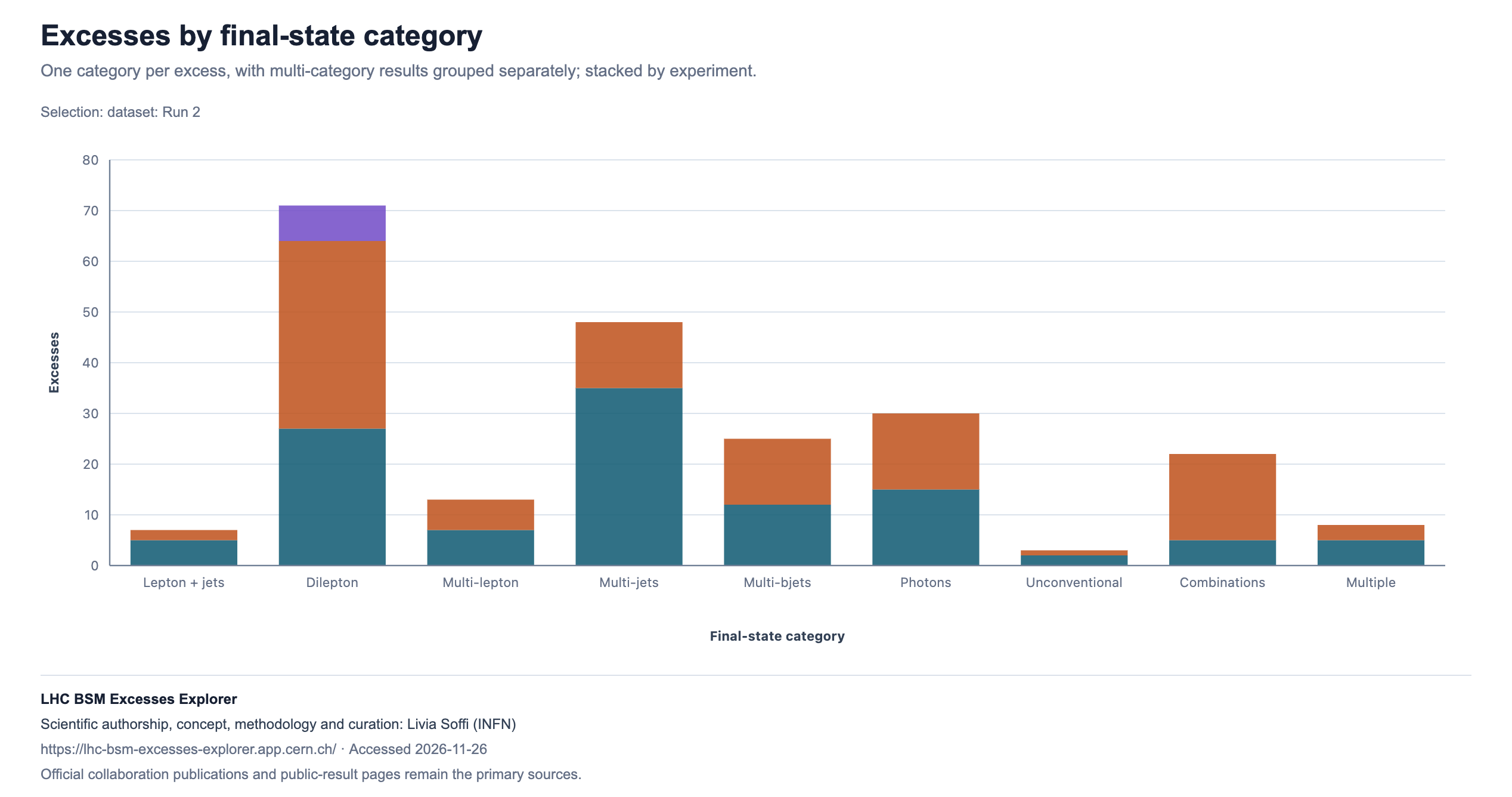}
  \caption{Examples of application exports. The top panel shows experiment-specific mass spectra and published significances for entries with a single mass coordinate. The bottom panel shows counts by final-state category for the Run~2 selection. Each bar counts records in the curated inventory.}
  \label{fig:explorer-views}
\end{figure}

The cascade \(X\to YH\to b\bar b\gamma\gamma\) illustrates how the catalogue links a benchmark to relevant searches. Here \(b\) and \(\bar b\) denote a bottom quark and antiquark, respectively, and \(\gamma\) denotes a photon. The ATLAS Run~2 maximum occurs at \((575,200)\,\mathrm{GeV}\), whereas the CMS maximum occurs at \((650,90)\,\mathrm{GeV}\)~\cite{atlascascade,cmscascade}. The later ATLAS Run~2+3 analysis covers the CMS benchmark~\cite{atlasupdate}. A benchmark-level comparison uses the result at that mass tuple under compatible production and decay assumptions. Searching for \texttt{HIG-21-011} in the \emph{Reference} field retrieves the source result and the reviewed ATLAS references (Fig.~\ref{fig:explorer-table}). The linked documents are the starting point for checking the common benchmark.

The spectrum and final-state views illustrate two projections of the same inventory (Fig.~\ref{fig:explorer-views}). The spectrum requires a single reported mass coordinate. The final-state view also includes entries described by non-mass observables, so it can be used to inspect the composition of a filtered selection.

\clearpage
\section{Catalogue scope and outlook}
The application provides an entry point for surveying reported BSM excesses and finding the public results needed for follow-up. Its coverage is defined by the source catalogues and the recorded selection and review rules. Counts describe the curated snapshot for the documented set of searches and selection criteria. Assessment at a common benchmark requires comparable signal hypotheses and reported quantities, such as limits on the signal rate. A joint statistical interpretation additionally requires the relevant likelihood information and a treatment of correlations. The application and dated snapshot identify the derived inventory. The linked publications remain the sources for the experimental results.

The author plans to rerun the documented workflow every two to three months using the latest public catalogues, recording the results added or updated in each cycle. The counterpart review can then be extended to newly available searches and to additional entries in the inventory. The application could also incorporate dedicated plots and distributions for specific signal models, following the extended Higgs-sector survey in showing search coverage and reported deviations in the relevant parameter space. Dated, citable releases of the combined BSM workbook would allow readers to reproduce the selections and plots associated with a given release. The author foresees future collaboration with the LHC BSM WG to guide these updates and refine how common benchmarks and counterpart relations are documented.

\section*{Acknowledgements}
The author thanks the LHC collaborations, especially ATLAS, CMS and LHCb, for their public results and documentation, CERN for hosting the application, and the LHC BSM Working Group and its Steering Committee for fostering exchanges between experimentalists and theorists. The author thanks colleagues across the experiments and the theory community for fruitful discussions, feedback and suggestions on cross-experiment comparisons.

\end{document}